\documentclass[12pt]{iopart}
\usepackage{harvard}
\usepackage[table]{xcolor}
\usepackage{graphicx}
\usepackage{multirow}
\usepackage{hyperref}
\usepackage{arydshln}
\usepackage{soul}
\usepackage{lineno}

\newcommand{\Bcite}[1]{{\color{blue}\cite{#1}}}
\newcommand{\Bref}[1] {{\color{blue}\ref{#1}}}

\begin{document}
\title[Influence of MRI segmentation accuracy on EF for TMS and tES]{Influence of segmentation accuracy in structural MR head scans on electric field computation for TMS and tES}
\author{Essam A Rashed$^{1,2}$, Jose Gomez-Tames$^{1,3}$, Akimasa Hirata$^{1,3}$}
\address{$^1$Department of Electrical and Mechanical Engineering, Nagoya Institute of Technology, Nagoya 466-8555, Japan}
\address{$^2$Department of Mathematics, Faculty of Science, Suez Canal University, Ismailia 41522, Egypt}
\address{$^3$Center of Biomedical Physics and Information Technology, Nagoya Institute of Technology, Nagoya 466-8555, Japan}
\ead{essam.rashed@nitech.ac.jp}
\vspace{10pt}


\begin{abstract}

In several diagnosis and therapy procedures based on electrostimulation effect, the internal physical quantity related to the stimulation is the induced electric field. To estimate the induced electric field in an individual human model, the segmentation of anatomical imaging, such as magnetic resonance image (MRI) scans, of the corresponding body parts into tissues is required. Then, electrical properties associated with different annotated tissues are assigned to the digital model to generate a volume conductor. However, the segmentation of different tissues is a tedious task with several associated challenges specially with tissues appear in limited regions and/or low-contrast in anatomical images. An open question is how segmentation accuracy of different tissues would influence the distribution of the induced electric field. In this study, we applied parametric segmentation of different tissues to exploit the segmentation of available MRI to generate different quality of head models using deep learning neural network architecture, named ForkNet. Then, the induced electric field are compared to assess the effect of model segmentation variations. Computational results indicate that the influence of segmentation error is tissue-dependent. In brain, sensitivity to segmentation accuracy is relatively high in cerebrospinal fluid (CSF), moderate in gray matter (GM) and low in white matter for transcranial magnetic stimulation (TMS) and transcranial electrical stimulation (tES). A CSF segmentation accuracy reduction of 10\% in terms of Dice coefficient (DC) lead to decrease up to 4\% in normalized induced electric field in both applications. However, a GM segmentation accuracy reduction of 5.6\% DC leads to increase of normalized induced electric field up to 6\%. Opposite trend of electric field variation was found between CSF and GM for both TMS and tES. The finding obtained here would be useful to quantify potential uncertainty of computational results.
\end{abstract}

\vspace{2pc}
\noindent{Keywords}: Image segmentation, MRI, ForkNet, uncertainty analysis, TMS, tES

\citationmode{abbr}

\section{Introduction}

In non-invasive electrostimulation of the brain, different methods have been proposed and used for neuroscience research and clinical applications~\Bcite{Rossini2015CN,Miniussi2013NBR}. Commonly used methods are transcranial electrical stimulation (tES) with direct/alternating current \Bcite{Paulus2011NR} and transcranial magnetic stimulation (TMS)~\Bcite{Rosen2009CPHR}. Although TMS has been approved as a clinical procedure for the treatment of neurological disorders (such as depression), several questions are still open for personalized TMS simulation~\Bcite{Valero2017NBR}. In TMS, pulse whose center frequency is the order of kHz is used. In tES, the frequency widely used is from 0 to 100 Hz, in addition that kHz range is used for interferential stimulation. Electrostimulation is established effect for electromagnetic field exposures. According to the international guidelines/standard for human protection from electromagnetic fields, the upper frequency where the stimulation is dominant is 5-10 MHz~\Bcite{IEEEC9512019,ICNIRP2020}. At frequencies lower than 300-400 Hz, the synaptic effect is dominant whereas axonal stimulation would be dominant at higher frequencies (up to 5-10 MHz).

Even for different types of stimulation mechanism, the common physical measure to estimate the electrostimulation effects is the induced electric field within the target tissues/areas. To compute the induced electric field, a heterogeneous digital model generated from anatomical images is required for simulation studies~\Bcite{GomezTames2020PMBreview,Bikson2012CEN}. These models are commonly generated through the segmentation of anatomical images such as MRI to represent different anatomical structures~\Bcite{Huang2013JNE,Datta2011BS}. Segmentation can either be performed manually, which is a time-consuming process or automatically with potential less accuracy. There are different scales where the compromise always exists between computational efforts/time and segmentation accuracy~\Bcite{Baxter2018MIA}. Moreover, segmentation quality is still biased even if it is conducted manually by experts. Digitization and segmentation for images of biological tissues are known to suffer from partial volume effect (PVE), where some digital voxels may contain information of several tissues especially within the border regions~\Bcite{Ballester2002Media}. 

Enabling precision brain stimulation requires accurate annotation of different head anatomy and exact brain mapping that can be generated within clinically reasonable time~\Bcite{Windhoff2013HBM}. Therefore, fast and accurate segmentation would lead to a more feasible personalized brain stimulation. Within this scope, different methods are used to perform automatic segmentation of the brain~\Bcite{despotovic2015MRI}, but only a few attempts exist for the segmentation of all head tissues~\Bcite{Makris2008MBEC,Laakso2015BS,Rashed2019NI,Huang2019JNE,Penny2011,Nielsen2018NI,Thielscher2015EMBC}. While brain tissues are the main focus of this problem, non-brain tissues are also important to be identified correctly as it has non-negligible influence on the computation of induced electric field in particular for tES. Inappropriate modeling of non-brain tissues may lead to the incorrect distribution of electric field in the brain~\Bcite{Lee2018CNe,Janssen2013PMB,Thielscher2011NI,Opitz2011NI,Miranda2003TBE}.

More recently, deep learning approaches are emerging as the leading segmentation strategy with ability to generate a human-level accuracy in short time~\Bcite{Wachinger2018NI,Rashed2019NI,Henschel2020NI}. Unlike conventional automatic segmentation, deep learning-based segmentation is a powerful approach because it can easily learn, observe and extract anatomical features without pre-engineered feature design~\Bcite{Akkus2017JDI}. A common trend in segmentation validation is to compare different methods with a golden truth that is likely offered through manual annotation by experts. However, it is still unclear how much accuracy is required for potentially accurate electromagnetic computations. Considering electromagnetic brain stimulation, do we really need to have a very accurate segmentation? There is no explicit answer to this question considering the use of deep learning automatic segmentation. In this context, it is a compromise between the segmentation accuracy, that would lead to more realistic distribution of electric field, and segmentation speed, that is likely to improve the applicability and feasibility of clinical use. Therefore, it is important to understand the influence of deep learning multi-tissue segmentation accuracy of anatomical MRI scans. Up to the best of authors' knowledge, however, no study has evaluated this issue.  

In this study, we apply a deep learning-enabled parametric segmentation of T1-weighted MRI scans to generate different head models with different quality. Then, segmented head models are used to evaluate the induced electric field for typical scenarios of TMS and tES. 



\section{Materials and methods}

\subsection{Structural MRI data}

A set of T1-weighted structural MRI scans (256$\times$256$\times$256~voxels) with resolution of 1.0$\times$1.0$\times$1.0~mm was obtained from freely available dataset (NAMIC: Brain Multimodality~\footnote{\color{blue}\it https://www.insight-journal.org/midas/collection/view/190}). The number of selected subjects are 18 and each subject is defined as a combination of 13 different tissues (namely: blood, bone (cortical), bone (cancellous), cerebellum, cerebrospinal fluid, dura, fat, grey matter, mucous tissue, muscle, skin, vitreous humor, and white matter). All MRI volumes are normalized with zero mean and unit variance, followed by scaling in the range of $[0,1]$. Generation of tissue-based probabilistic maps is conducted using our deep learning architecture ForkNet~\footnote{\color{blue}\it Open source code avaibale at:~https://github.com/erashed/ForkNet}~\Bcite{Rashed2019NI}. 


\begin{figure*}
\centering
\includegraphics[width=1.0\textwidth]{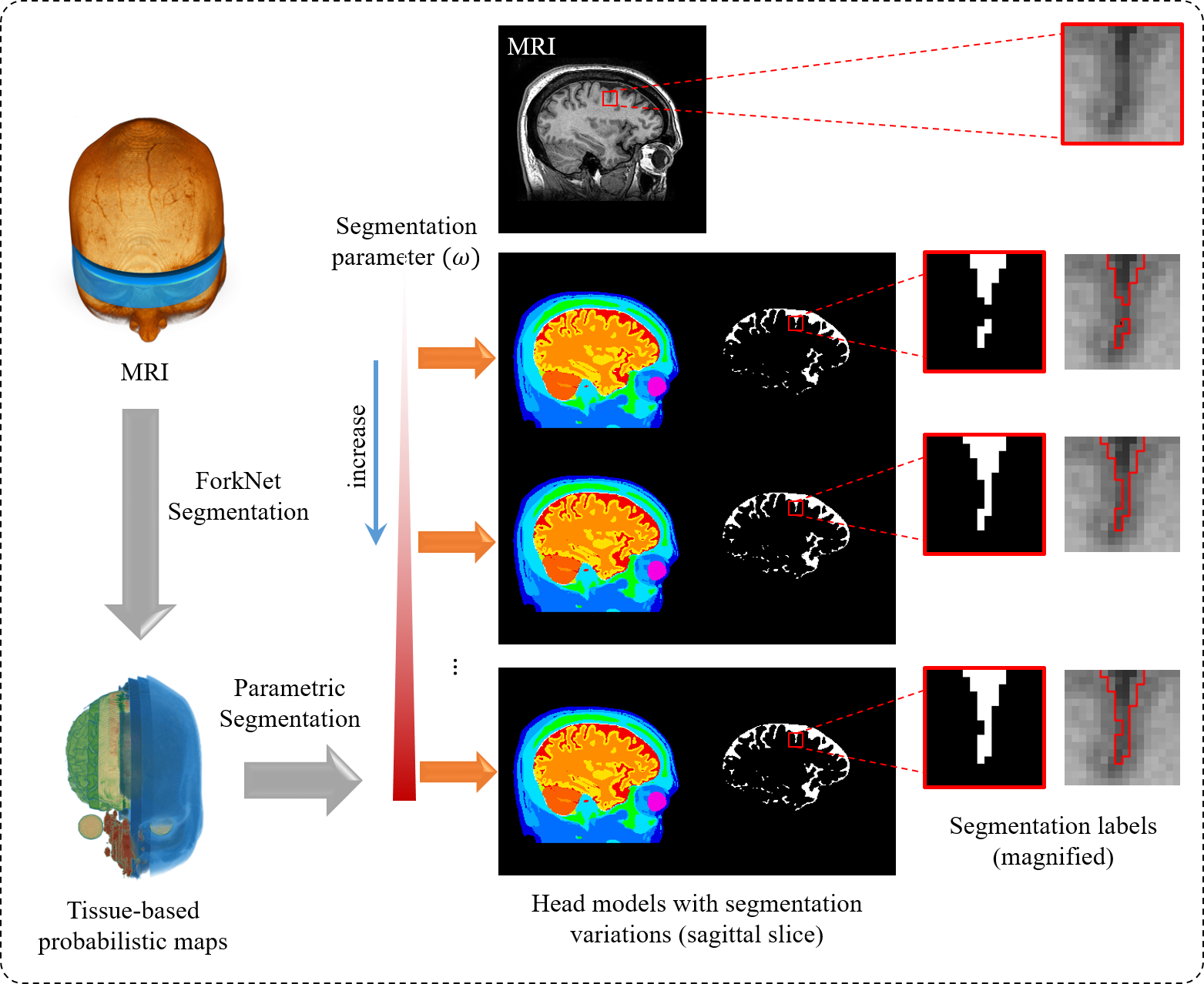}
\caption{Personalized head model is generated through automatic segmentation using ForkNet to generate tissue-based probabilistic maps. Segmentation of different tissues are fine tuned using parametric segmentation that emphasis a single tissue with weighting parameter $\omega$. In this example, cerebrospinal fluid (CSF) is emphasized and segmentation variations can be recognized in magnified binary regions (white is CSF and black is other tissues).}
\label{model}
\end{figure*}

\subsection{Parametric segmentation using ForkNet}

In our previous study, ForkNet was used to generate personalized head models directly from structural MRI~\Bcite{Rashed2019NI}. The network input is T1-weighted MRI and outputs are a tissue-based probability map. ForkNet is based on T1 rather than T2 and thus high-contents tissue is rather stable. Although there are several other methods that can be used for parametric segmentation, ForkNet is selected as it provide high quality segmentation of brain and non-brain tissues in short time. Let $M$ be the MRI volume, then ForkNet output is computed as:

\begin{equation}
L_{k,n}=\textnormal{ForkNet}(M_k), ~k=1,\dots,K,~ n=1,\dots,N,
\end{equation}
where $M_k$ is a 2D MRI slice and $L_{k,n}$ is the corresponding probability map of tissue $n$. Therefore, the reference head model is computed using the following SoftMax rule:

\begin{equation}
R_k(i,j)=\arg \max_{n} L_{k,n} (i,j), \forall i,j,k,
\label{softmax}
\end{equation}
which means that all tissues are treated equally and a slice pixel $R_k(i,j)$ is assigned to the tissue label $n$ that has the highest probability score. Probability maps generated using deep learning can be fine-tuned to generate different segmentation patterns by favor specific tissue distributions. This technique characterizes the segmentation variations within different automatic/manual segmentation frameworks. In some cases, especially within the border regions, the probability of single tissue is slightly higher than other potential ones, which demonstrate high uncertainty around these regions. Therefore, we have modified equation~(\Bref{softmax}) to demonstrate a tissue-based parametric segmentation defined as follows:

\begin{equation}
R_k^{(\tilde{n},\omega)}(i,j) = \arg \max_n \left[ (1+\lambda^{(\tilde{n},\omega)}) L_{k,n}(i,j) \right],
\label{softmax2}
\end{equation}
\begin{equation}
\lambda^{(\tilde{n},\omega)}=\left\{
\begin{array}{ll}
\omega & n=\tilde{n} \\
0 & n \neq \tilde{n}\\
\end{array},
\right.
\end{equation}
where $\tilde{n}$ is the label corresponding to the emphesised tissue with weighting factor $\omega$. When $\omega=0$, equation~(\Bref{softmax2}) is equivalent to equation~(\Bref{softmax}), while increasing $\omega$ will give a higher segmentation favor for the corresponding tissue $\tilde{n}$ over surrounding tissues as shown in figure~\Bref{model}. By this way, we can generate several segmentation versions of the same subject that simulate the variability caused by segmentation uncertainty. We have limited the change in each version to a single tissue only to clearly evaluate the effect of segmentation variations within a single tissue, though generalization to more than single tissue is direct. To demonstrate the effect of parametric segmentation using equation~(\Bref{softmax2}), we address the construction of several versions of head models using different values of $\tilde{n}$ and $\omega$.


\subsection{TMS simulation}

The generated head models with different segmentation are used to compute the brain induced electric field considering TMS simulations. A figure-eight magnetic stimulation coil with outer and inner diameters of 97 $mm$ and 47 $mm$, respectively is modeled with thin-wire approximation. The magnetic vector potential is computed using the Biot-Savart law for the coil located over the scalp to target the hand motor area of the brain. The induced electric field is determined from the vector potential using the scalar potential finite difference by assuming the magneto-quasi-static approximation~\Bcite{Barchanski2005PMB,Plonsey1967BMB,Hirata2013PMB}. Given the vector potential ($A_0$), we compute the scalar potential through solving the following equation:
\begin{equation}
\nabla [\sigma (-\nabla \psi -j w A_0)] =0,
\end{equation}
where $\sigma$ and $w$ are the tissue uniform conductivity and angular frequency, respectively. The tissue conductivity is assumed to be isotropic. and is computed using a fourth order Cole-Cole model with at frequency of 10 kHz, as detailed in~\Bcite{Gabriel1996PMBII}. Specifically, we have used the same tissue conductivity values listed in our previous study~\Bcite{Rashed2020TMI}. Finally, the induced electric field $E$ is calculated from
\begin{equation}
E=-\nabla \psi - \frac{\partial}{\partial t}A_0.
\end{equation}


\subsection{tES simulation}

The tES simulation is designed using transcranial direct current stimulation (tDCS) scenario with two electrodes attached to the scalp. Two sets of 20 $mm$ (and 50 $mm$) electrodes attached to the C3-Fp2 positions (10-20 electroencephalogram system) with injected current of 2 mA. The electric potential produced by the current injection was computed using the scalar potential finite-difference (SPFD) method~\Bcite{Dawson1998TMag}. The computation is accelerated using the successive-over-relaxation and multi-grid methods~\Bcite{Laakso2012PMB} and tissue electrical conductivity values are the same as those used in~\Bcite{Rashed2020NN}. The SPFD method is used to solve the scalar-potential equation:

\begin{equation}
\nabla(\sigma \nabla \phi)=0,
\end{equation}
where $\phi$ and $\sigma$ are the scalar potential and tissue conductivity, respectively. We consider the maximum electric field strength in brain region M1 that corresponds to hand motor cortex.


\begin{figure*}
\centering
\includegraphics[width=1.0\textwidth]{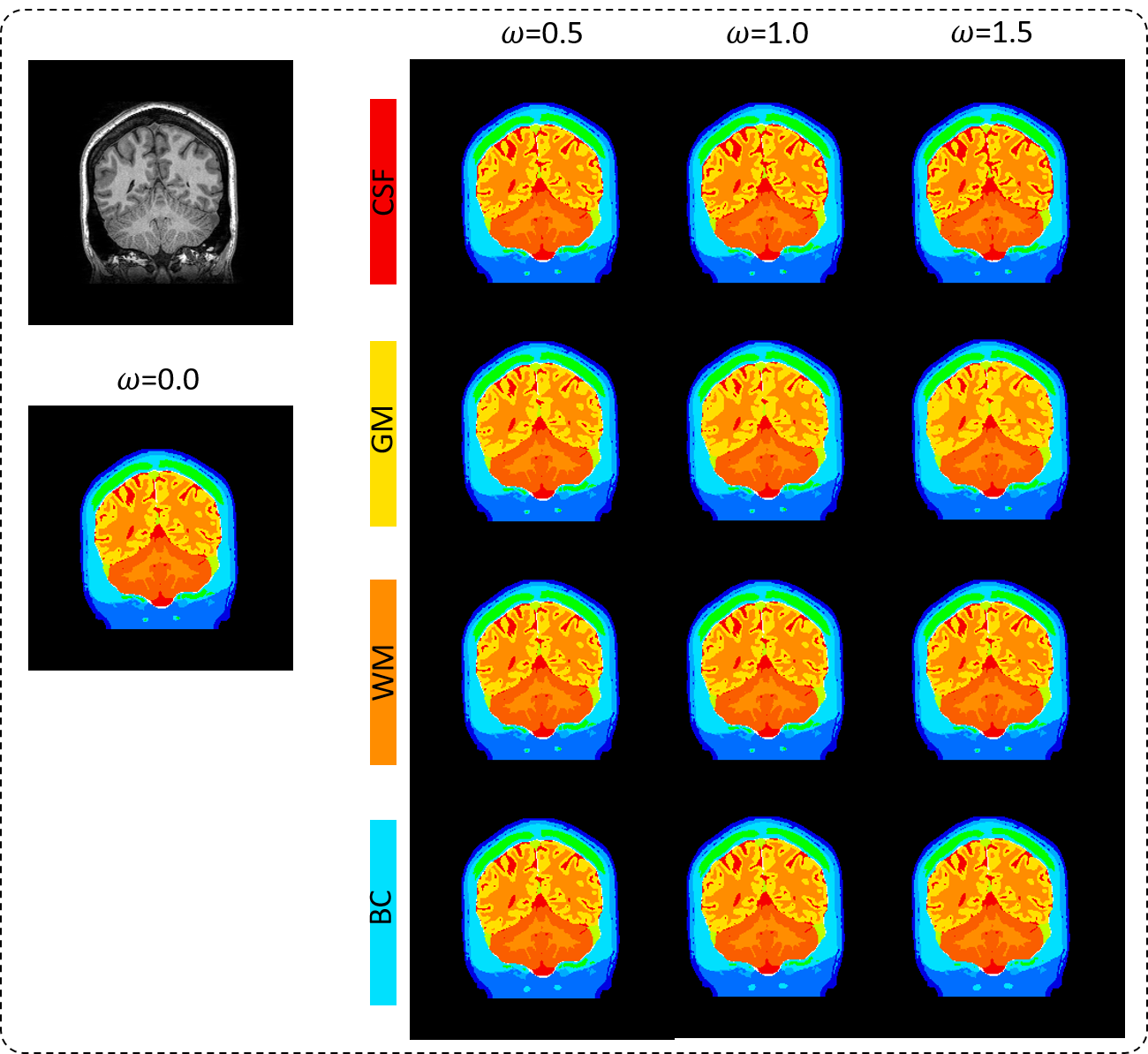}
\caption{Example of different head models generated using parametric segmentation. Left are MRI (T1) slice with standard model (no tissue emphasis). Right are head models with different tissue emphasis and variations of parameter $\omega$. Tissues represent CSF, gray matter (GM), white matter (WM) and bone cortical (BC). Magnified regions are shown in figure~\Bref{SegROI} for better demonstration of segmentation variations.}
\label{Seg}
\end{figure*}


\begin{figure*}
\centering
\includegraphics[width=1.0\textwidth]{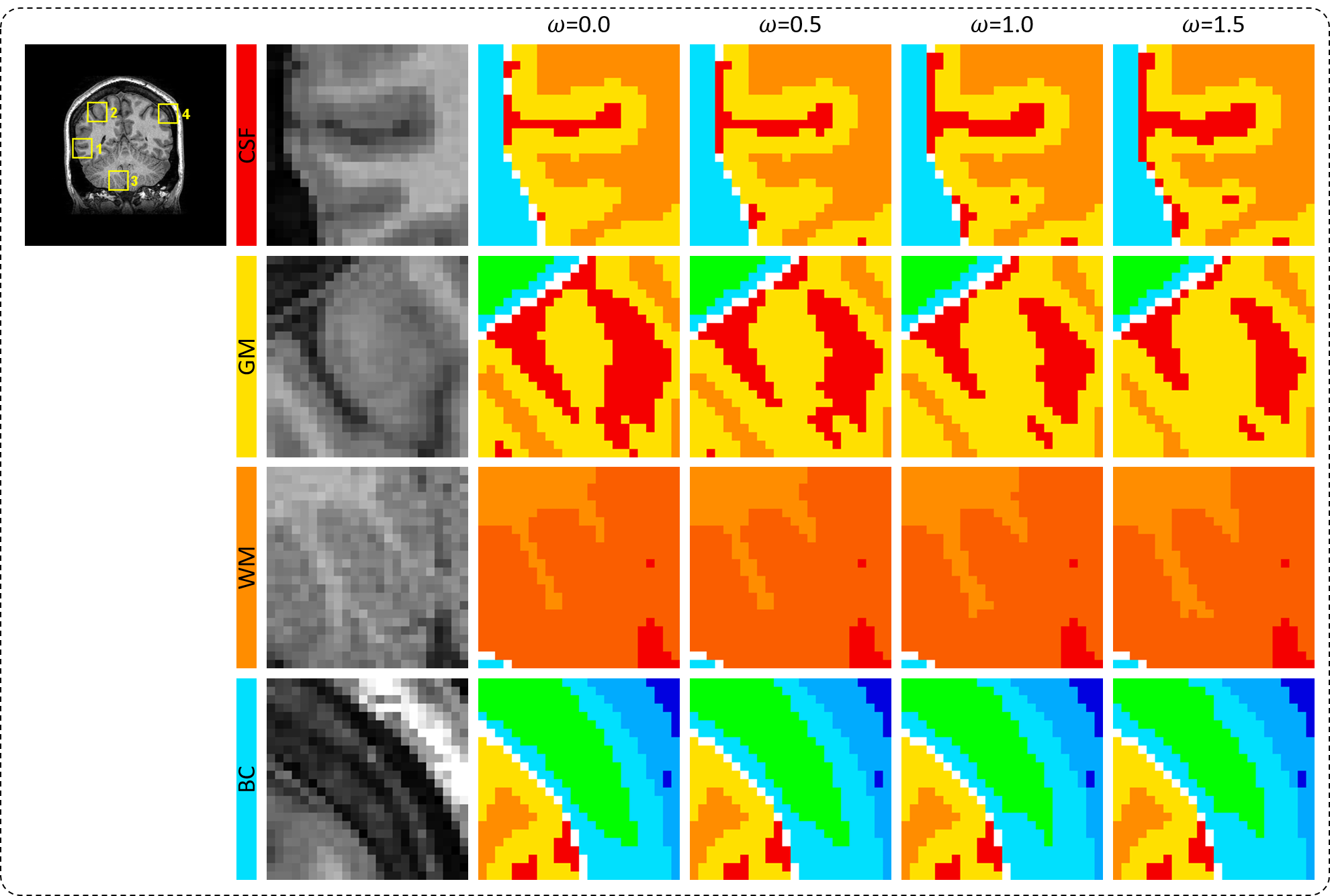}
\caption{Magnified regions of segmentation presented in figure~\Bref{Seg}. Left is MRI (T1) with ROI labeled as presented in order (from top to bottom). It is clear that increasing $\omega$ would lead to more favor segmentation score for emphasized tissue $\tilde{n}$.}
\label{SegROI}
\end{figure*}


\begin{figure*}
\centering
\includegraphics[width=1.0\textwidth]{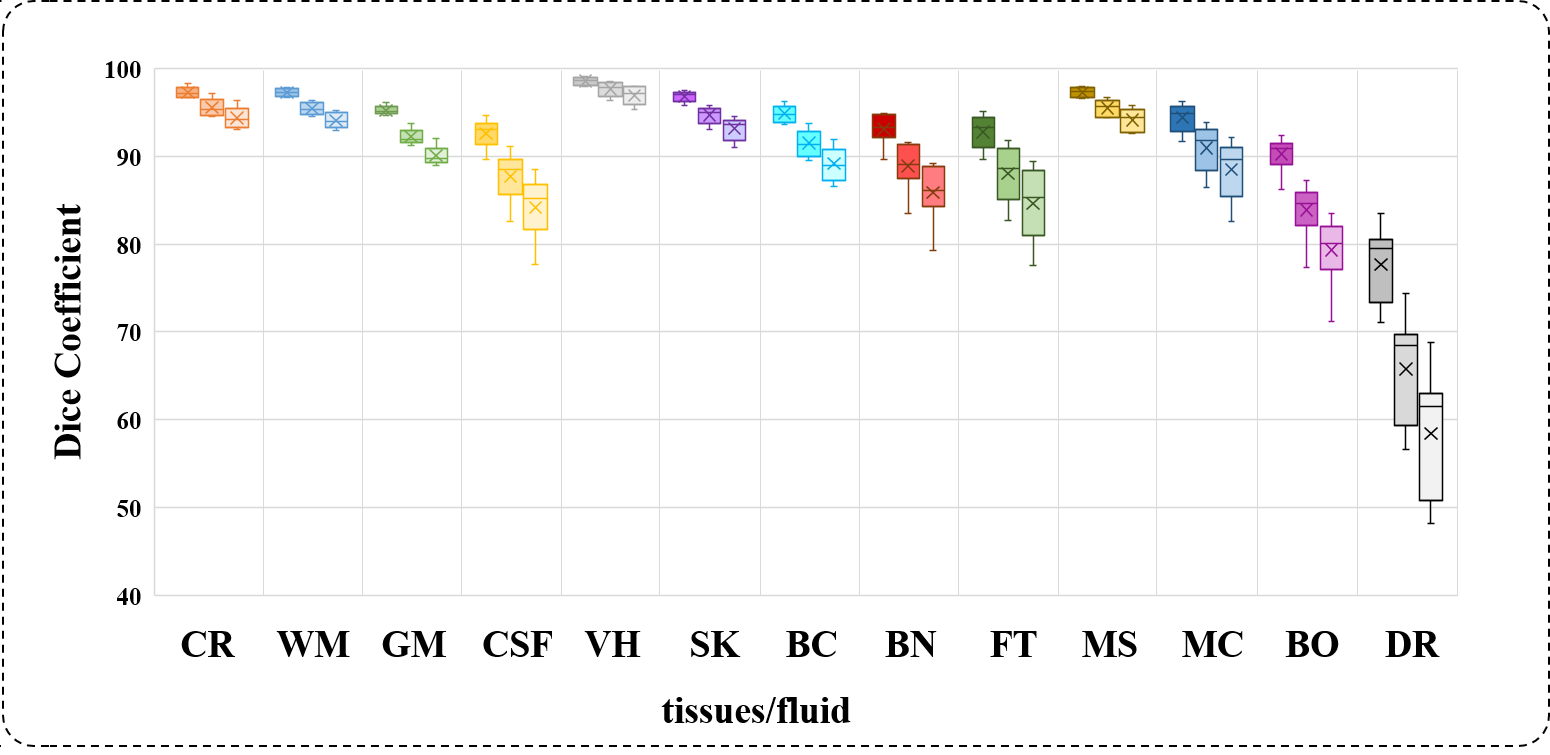}
\caption{Boxplots of DC computed from 18 subjects for different head tissues: cerebellum (CR), WM, GM, CSF, vitreous humor (VH), skin (SK), BC, bone cancellous (BN), fat (FT), muscle (MS), mucous tissue (MC), blood vessels (BO), and dura (DR). For each tissue, box plots represent $\omega=0.5, 1.0$ and $1.5$ are shown from left to right, respectively.}
\label{dc}
\end{figure*}

\section{Results}


\subsection{Tissue-based probabilistic maps}

The tissue probability maps $L_n$ for all 18 subjects are generated using ForkNet trained through leave-one-out cross-validation strategy. The network is trained by minimizing the cross-entropy cost function using ADAM algorithm~\Bcite{Kingma2014arXiv} with ground truth segmentation labels generated using a semi-automatic method detailed in~\Bcite{Laakso2015BS}. We consider 50 epochs with 2 image slices per batch. Parametric head models $R^{(\tilde{n},\omega)}$ are computed for the 13 head tissues with $\omega=0.5, 1.0,$ and $1.5$ along with references head models ($\omega=0.0$). These values are chosen such that the resulting segmentation variations is consistent with the error range reported in AI-based MRI head segmentation methods \Bcite{Bernal20196AIM}. Considering, variable volume and shape of different tissues that is also presented in different probability maps by ForkNet, it is challenging to setup a tissue-based $\omega$ values. Therefore, a consistent value is used for different tissues taking into account different variability in segmentation error as shown later. Considering 18 subjects, 13 different tissues, and values of $\omega$, a total of 720 head models are generated (18 models for $\omega=0$ and 18$\times$13$\times$3 models for $\omega>0$).

An example of the generated head models is shown in figure~\Bref{Seg} and magnified regions are shown in figure~\Bref{SegROI}. It is clear from these results that changing parameter $\omega$ would represent some segmentation variations that likely occurs even if manual segmentation is employed. To evaluate the segmentation variation within the new head models, we compute the Dice coefficient (DC) defined as:

\begin{equation}
DC_{n,\omega}=\frac{2|R_{n} \cap R^{(\tilde{n},\omega)}_{\tilde{n}}|}{|R_{n}  |+|R^{(\tilde{n},\omega)}_{\tilde{n}}|} . 100\%,
\end{equation} 
where $R^{(\tilde{n},\omega)}_{\tilde{n}}\equiv \{ R^{(\tilde{n},\omega)} | R^{(\tilde{n},\omega)}(i,j,k) =\tilde{n} ~ \forall i,j,k\}$. In other words, $DC_{n,\omega}$ demonstrate how the tissue $n$ generated with segmentation parameter $\omega$ is consistent with the corresponding tissue segmentation in the standard model $R$. Results of $DC_{n,\omega}$ values for the 18 subjects are shown in figure~\Bref{dc}. Due to the tissue variability considering distribution, volume, surface, surrounding tissues, and segmentation quality in probability maps, different $DC$ values are observed for each tissue. While a relatively small change ($<10\%)$ can be observed in most tissues, a remarkable change is recognized in blood vessels ($\approx 12\%$) and dura ($\approx 25\%$). The main reason can be the limited contrast of blood vessels in MRI and small thickness in dura. It is also reported that blood vessels and dura segmentation using ForkNet is of low-quality compared to other head tissues~\Bcite{Rashed2019NI}.


\begin{figure}[htb]
\centering
\includegraphics[width=1.0\textwidth]{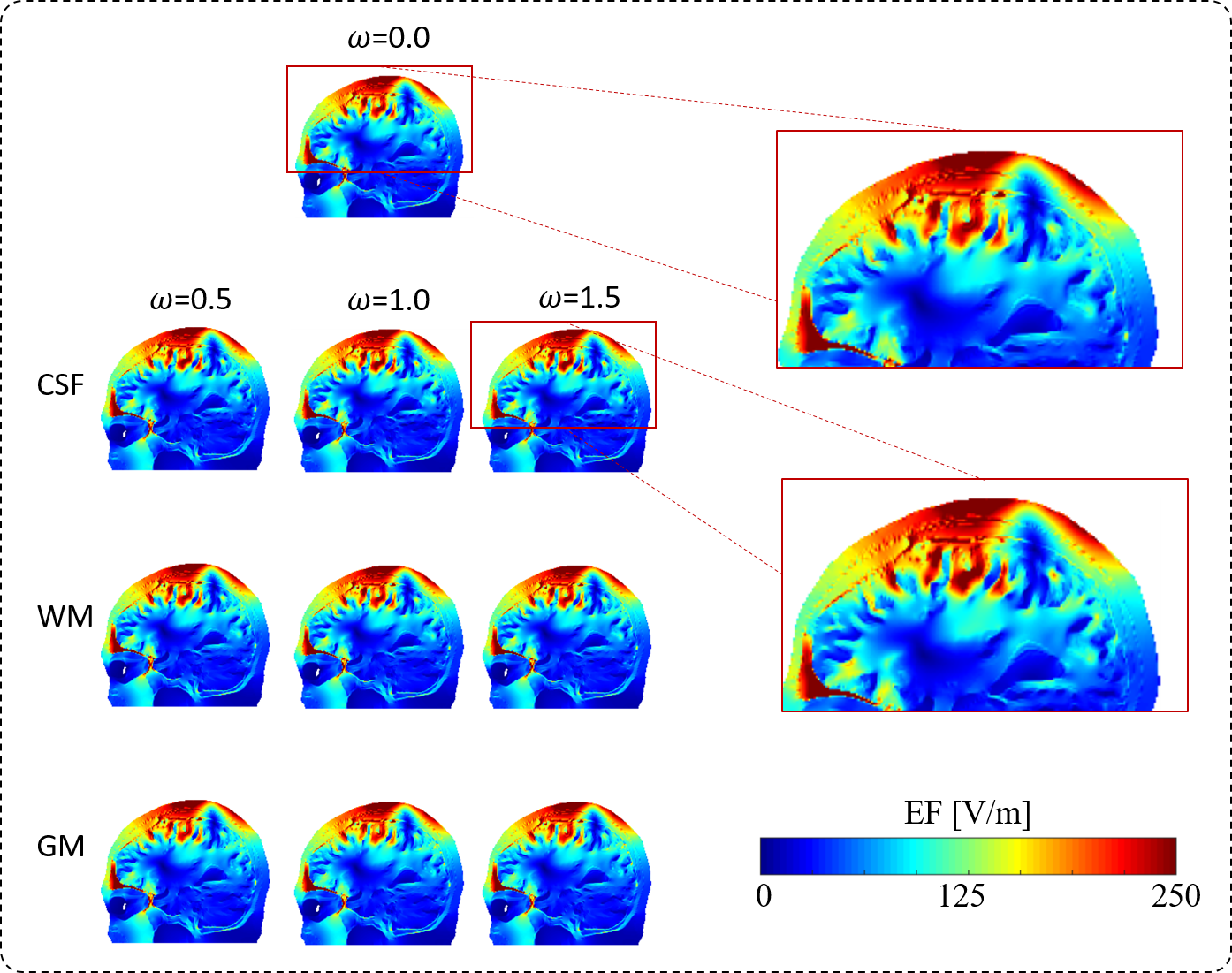}
\caption{Sagittal cross-section of induced electric field in models generated with variable segmentation of CSF, WM, and GM and the standard head model (subject: case01017). Right side demonstrate a magnified views.}
\label{tms17}
\end{figure}


\begin{figure}[htb]
\centering
\includegraphics[width=1.0\textwidth]{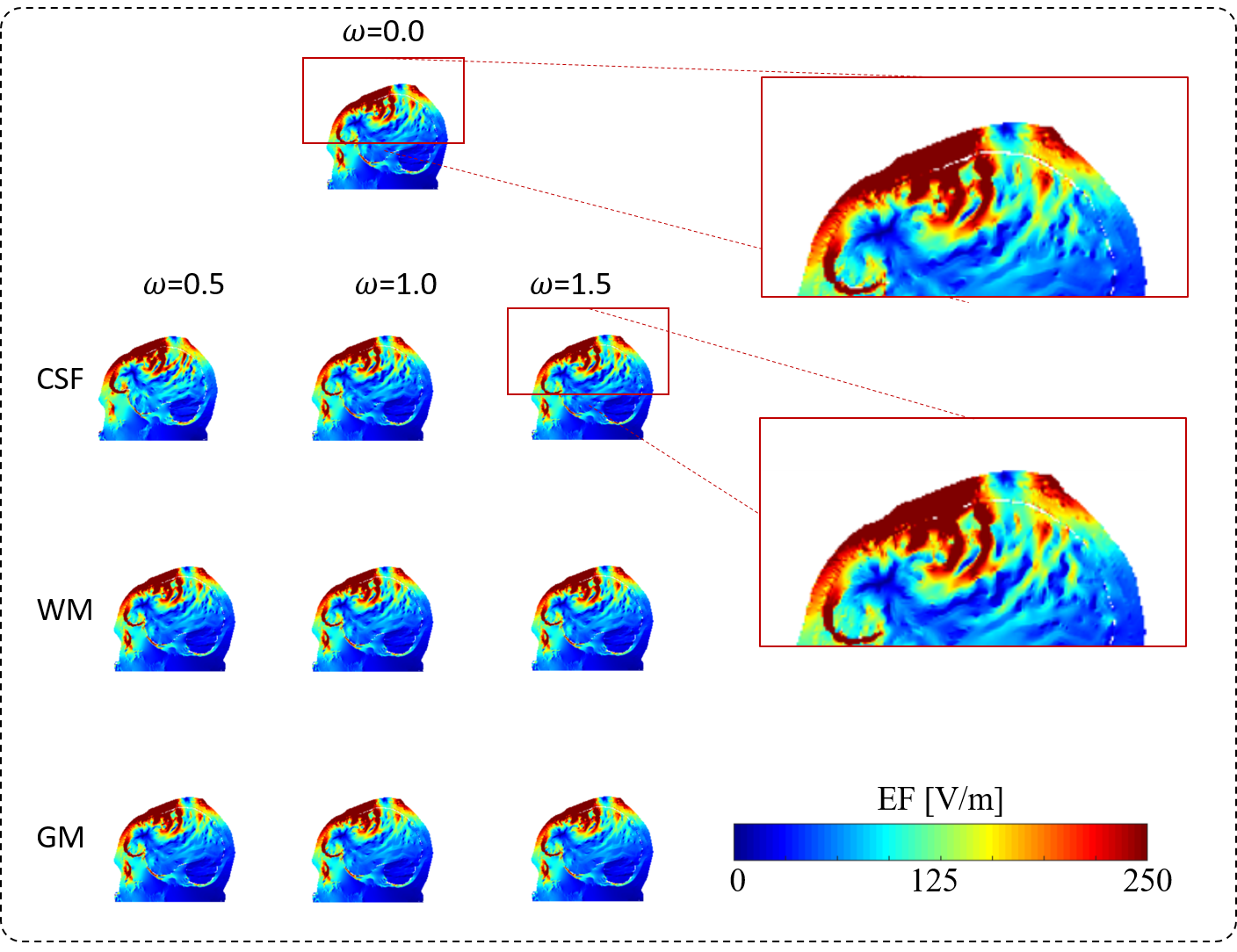}
\caption{Sagittal cross-section of induced electric field in models generated with variable segmentation of CSF, WM, and GM and the standard head model (subject: case01019).}
\label{tms19}
\end{figure}


\begin{figure}[htb]
\centering
\includegraphics[width=1.0\textwidth]{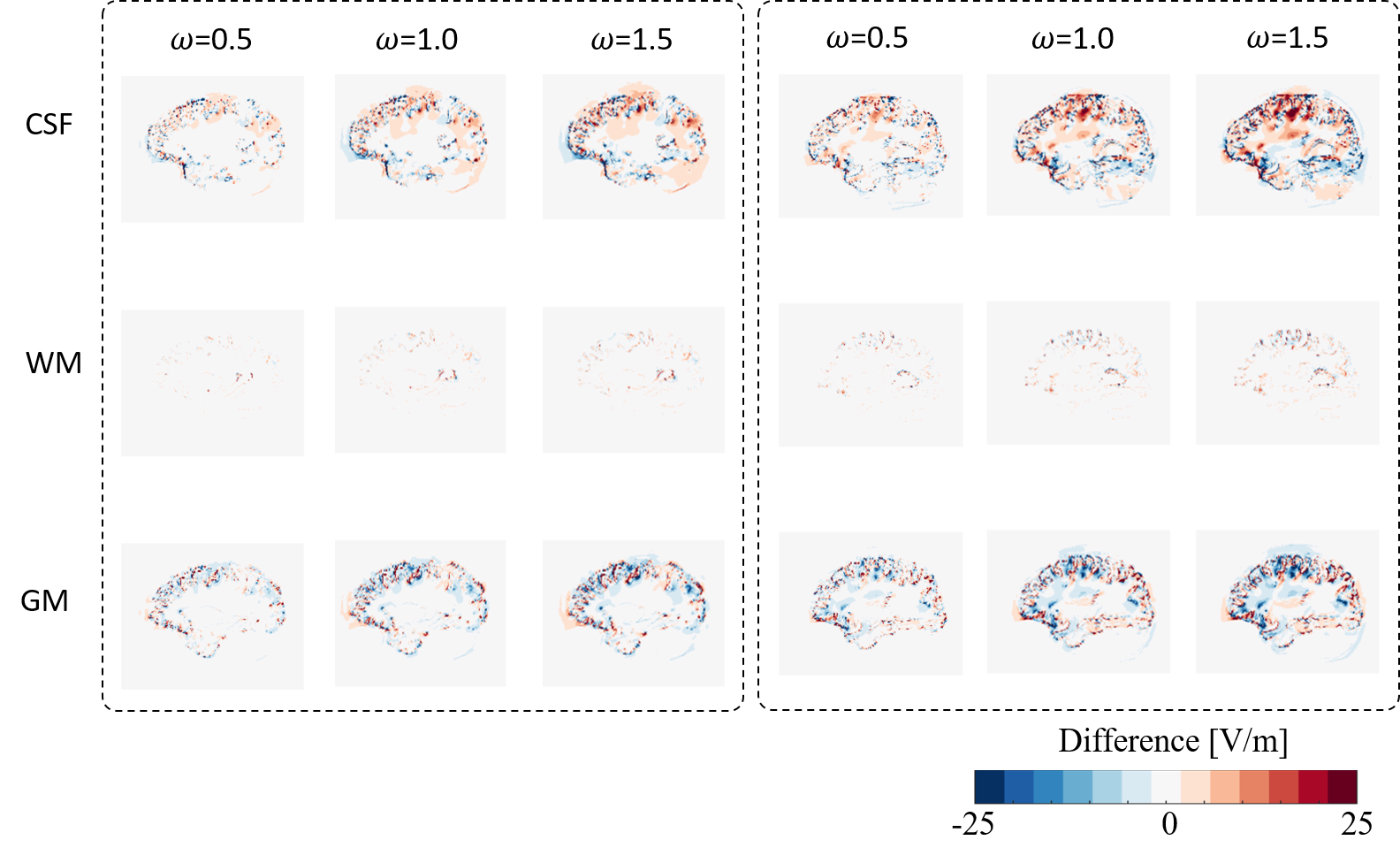}
\caption{Error maps of brain induced electric field shown in figures~\Bref{tms17} (left) and \Bref{tms19} (right). Significant change is recognized in CSF and GM compared to WM.}
\label{TMSerror}
\end{figure}


\begin{figure}[htb]
\centering
\includegraphics[width=1.0\textwidth]{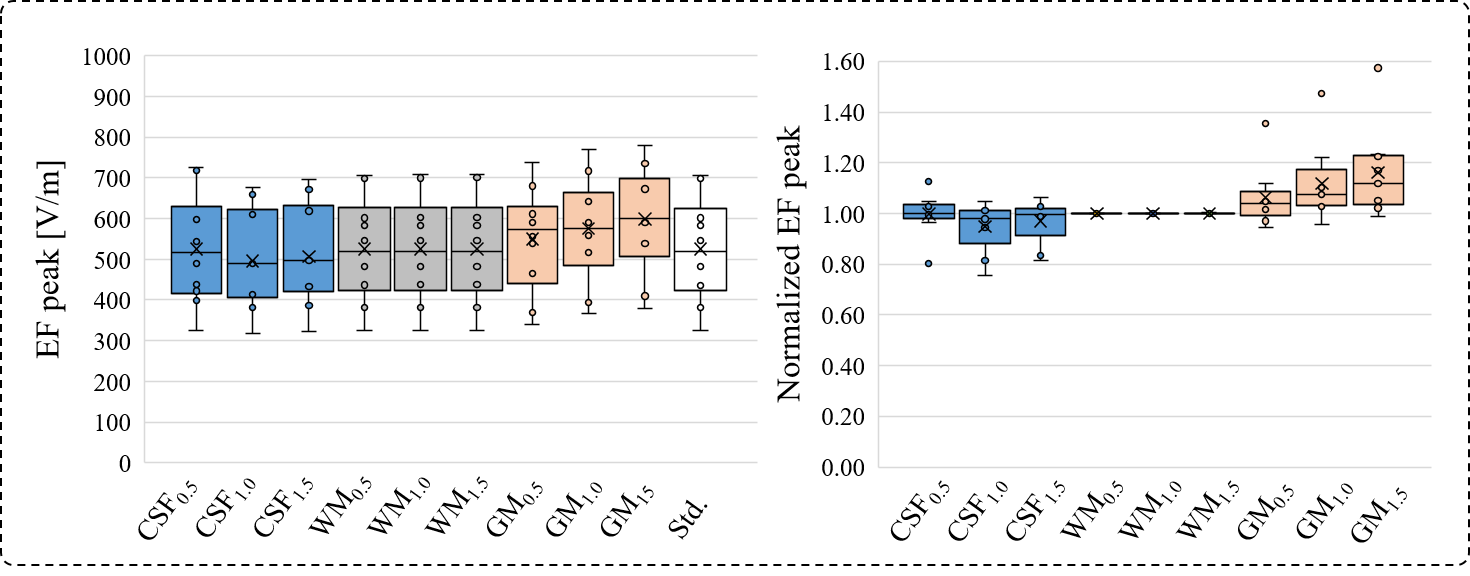}
\caption{Box plot of maximum field distribution in brain for different segmentation setups for TMS. Normalized values are the peak EF corresponsing to the peak EF in the standard model.}
\label{TMSplot}
\end{figure}


\begin{figure}[htb]
\centering
\includegraphics[width=1.0\textwidth]{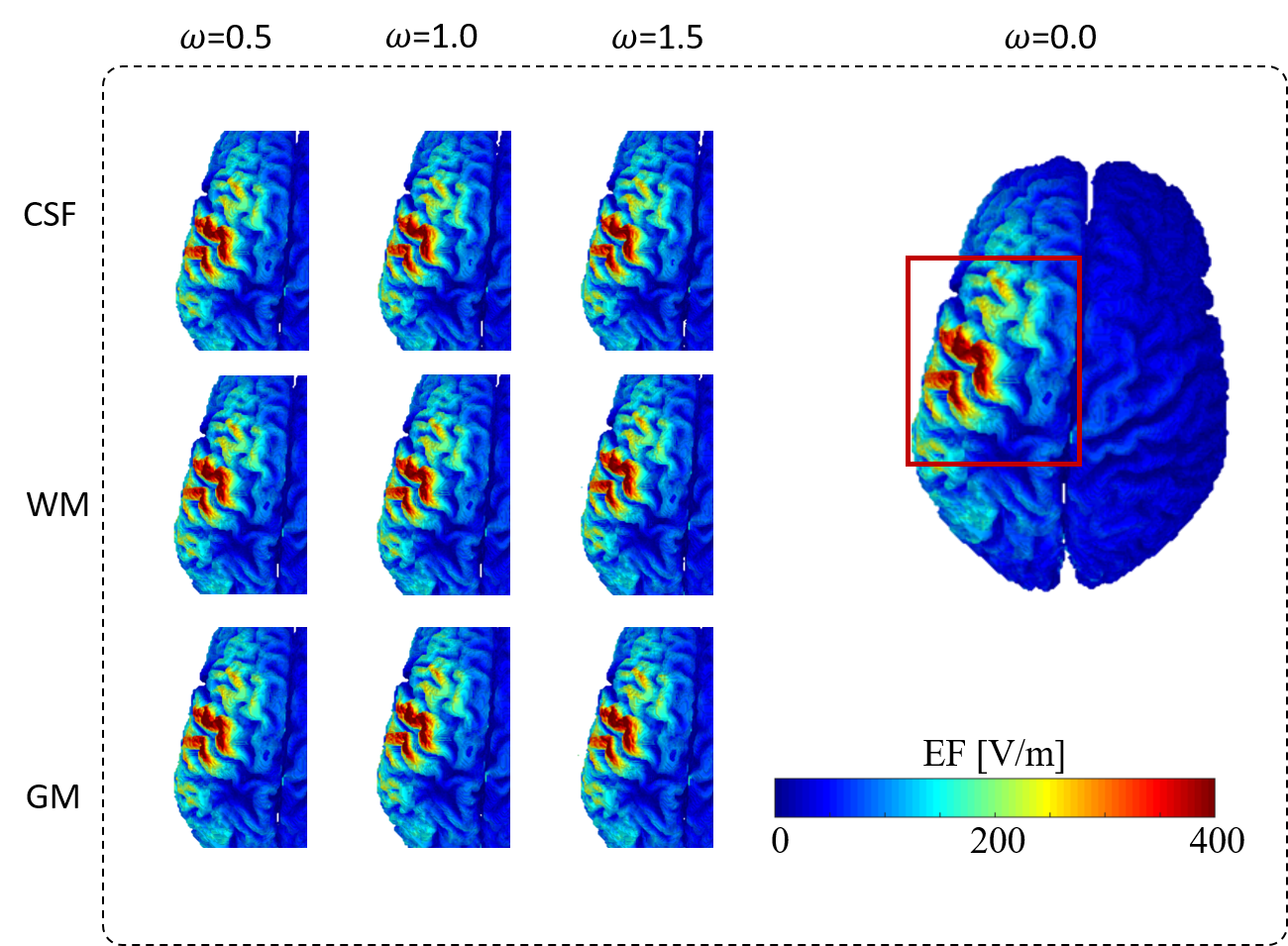}
\caption{TMS induced electric field map in brain (subject: case01028) with parametric segmentation emphasis CSF, WM and GM compared with values in standard segmentation.}
\label{tms28brain}
\end{figure}


\subsection{TMS experiment}

A set of randomly selected 10 head models are attributed to TMS simulation and the induced electric field is computed within the whole head. A sagittal cross-section electric field distribution in models with variable CSF, WM and GM segmentation along with standard model for two subjects are shown in figures~\Bref{tms17} and \Bref{tms19} with error maps in figure~\Bref{TMSerror}. The induced electric field within the brain cortical region of one subject with different segmentation setups is shown in figure~\Bref{tms28brain}. From these results, it is difficult to recognize a significant change in the electric field distribution. However, by looking at electric field maximum value in the hand knob region as shown in figure~\Bref{TMSplot}, we observed that a relatively higher variation in the metric for segmentation had the most impact on the electric field when compared with the reference model. In particular, the TMS-induced electric field was more sensitive to the segmentation variations of the CSF and GM. This is because the boundaries between tissues of different conductivity can strongly affect the field distribution in TMS. Also, the complex gyrification of GM makes that interface with CSF allows local hotspots of the induced electric field. On the other hand, variation of WM boundaries does not affect much the induced electric when observed in the cortical surface as in this work.


\begin{figure}[htb]
\centering
\includegraphics[width=1.0\textwidth]{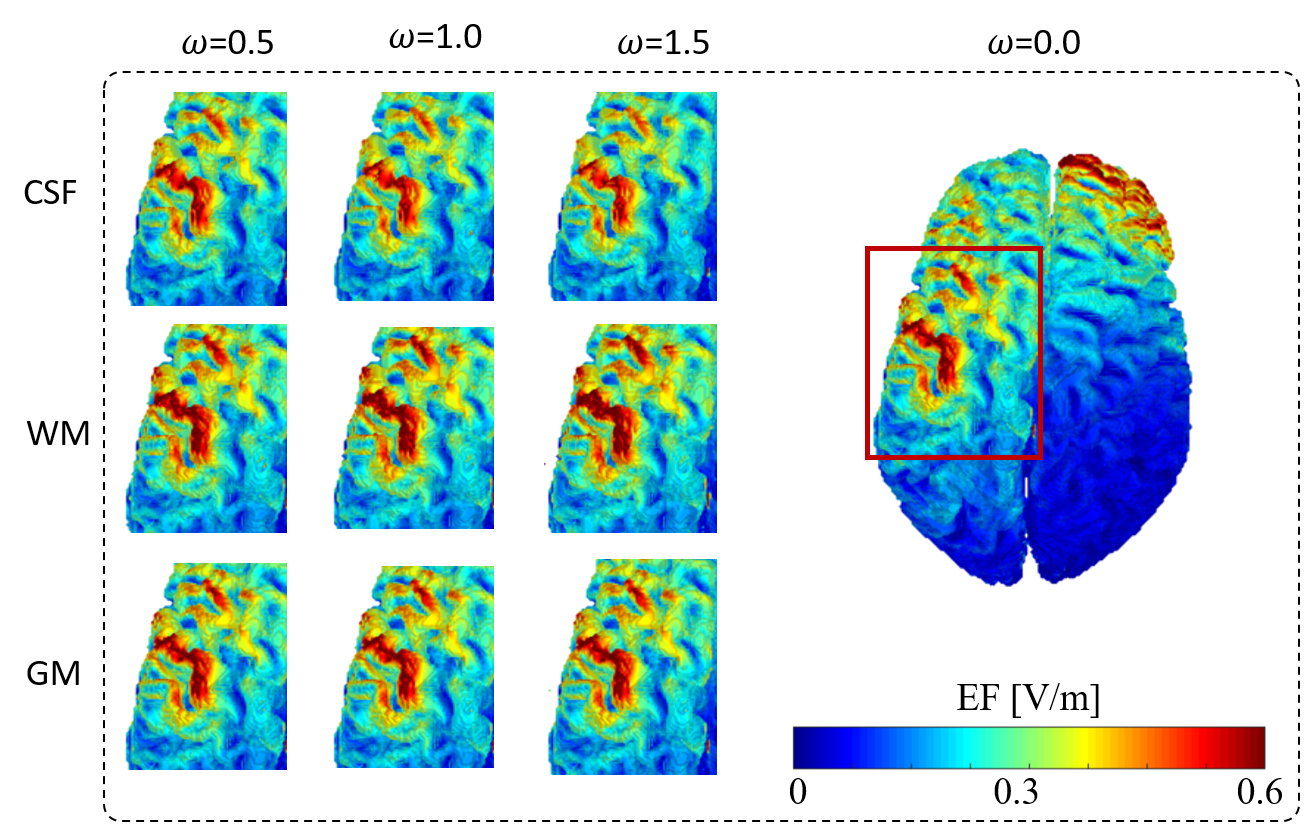}
\caption{Electric field distribution in the brain corresponding to tDCS with 20 $mm$ electrode with parametric segmentation for CSF, WM and GM along with standard segmentation (subject: case01028).}
\label{tDCS2mm}
\end{figure}


\begin{figure}[htb]
\centering
\includegraphics[width=1.0\textwidth]{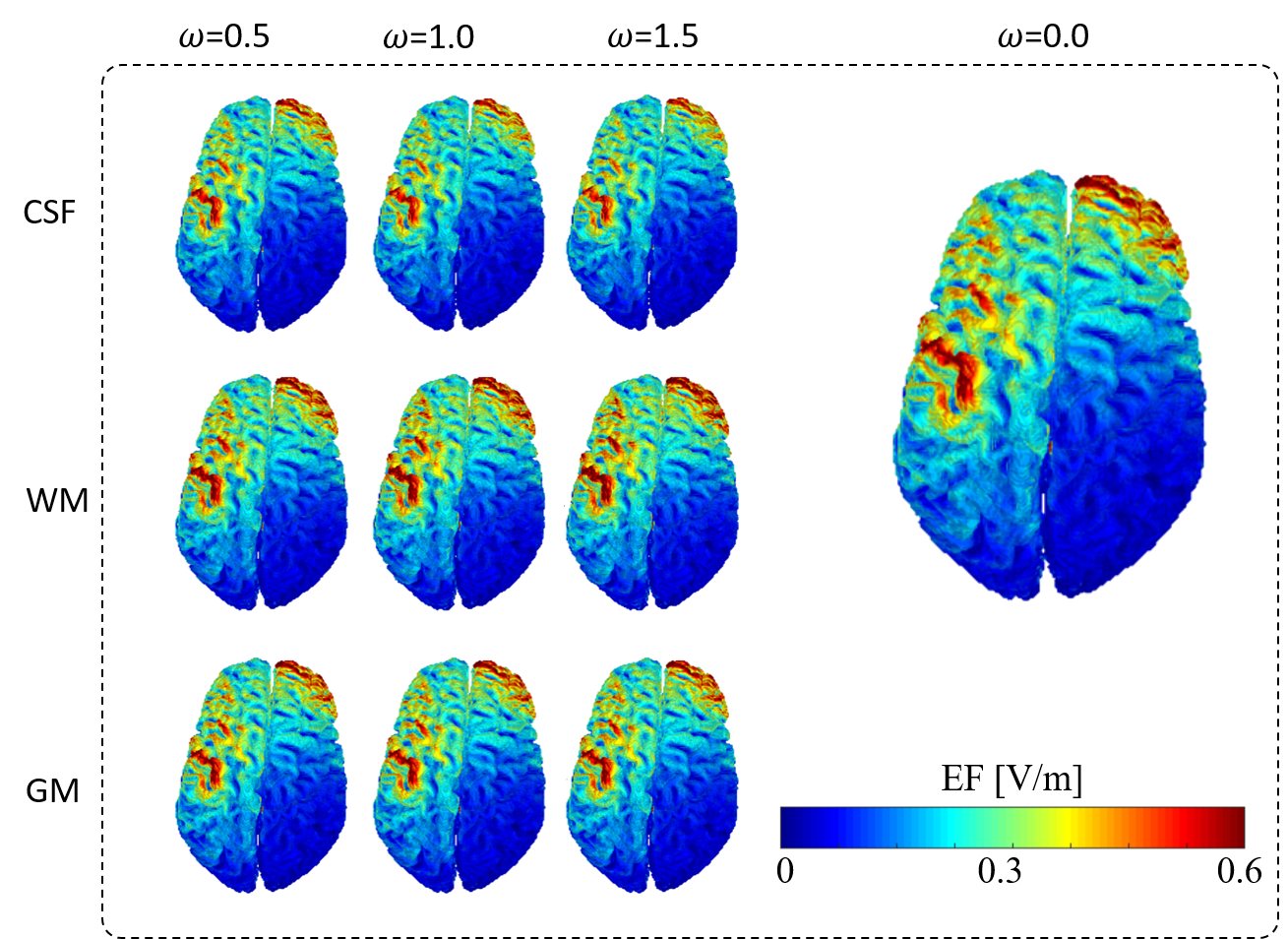}
\caption{Electric field distribution in the brain corresponding to tDCS with 50 $mm$ electrode with parametric segmentation for CSF, WM and GM along with standard segmentation (subject: case01028).}
\label{tDCS5mm}
\end{figure}


\subsection{tES experiment}

\begin{figure}[htb]
\centering
\includegraphics[width=1.0\textwidth]{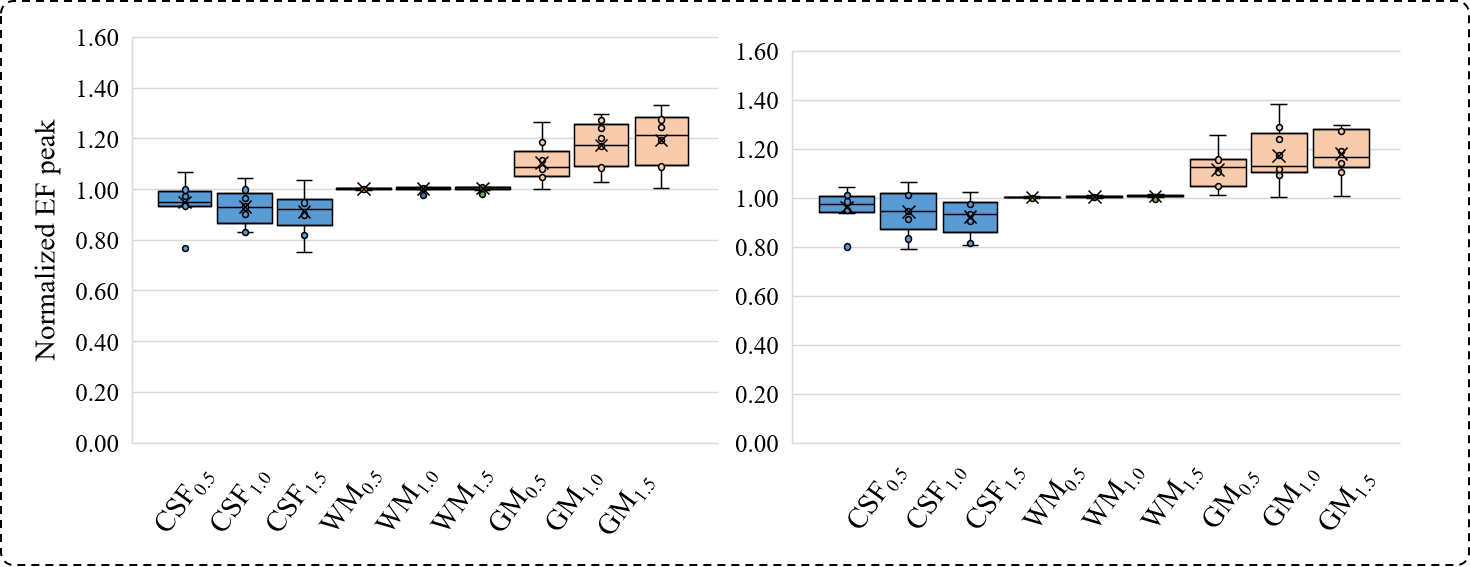}
\caption{Box plot of maximum field distribution in brain for different segmentation setups and tES with 20 $mm$ (left) and 50 $mm$ (right) size electrodes.}
\label{tDCSplot}
\end{figure}

The tES study was implemented using 10 subjects with parametric segmentation considering three tissues (CSF, WM and GM). We also keep the same $\omega$ values similar to the those used in the TMS study. A sample result for one subject with 20 $mm$- and 50 $mm$-size electrodes are shown in figures~\Bref{tDCS2mm} and \Bref{tDCS5mm}. A consistent tendency is observed for the segmentation variation of GM. Larger $\omega$ values lead to increase of the electric field. A marginal variation in electric field is recognized with the WM emphasized parametric segmentation. An opposite trend of the electric field distribution is observed as increasing parameter $\omega$ for CSF lead to reduction of the induced electric field. A box plot demonstrates the observed changes in the normalized peak electric field for 10 subjects is shown in figure~\Bref{tDCSplot}.


\begin{table*}
\centering
\footnotesize
\caption{Average (standard deviation) of the Euclidean distance of the hotspot location between segmented and standard head models (n = 10). The distance is zero in the case of WM variations.}
\label{Tab1}
\setlength{\tabcolsep}{3pt}
\begin{tabular}{|c|c|c|c|c|c|c|}
\hline
\multirow{2}{*}{Simulation} & \multicolumn{6}{c|}{Distance error (mm)} \\
\cline{2-7}
 & CSF$_{0.5}$ & CSF$_{1.0}$ & CSF$_{1.5}$ & GM$_{0.5}$ & GM$_{1.0}$ & GM$_{1.5}$\\
\hline 
TMS & 0.7 (0.8)& 1.0 (1.5)&1.5 (1.3)&1.9 (1.4)&1.8 (1.3)&2.0 (1.6)\\
\hline
tDCS & 0.7 (1.0)& 1.8 (1.1)& 2.0 (1.6)& 1.6 (1.4)& 1.8 (1.5)& 1.9 (1.5)\\
\hline  
\end{tabular}
\end{table*}

Away from the distributions of the induced electric field, it is also important to identify the location of maximum electric field intensity as the brain region that is most engaged in stimulation in different clinical applications. Therefore, we computed the Euclidean distance error to quantify the location change of the maximum in the motor area. The location of the maximum value for each segmented model is compared with the one in the standard. We confirmed shifted positions of the maximum in segmented models with different CSF and GM. There was no difference in the location for variations of WM. In the case of TMS, the maximum distance error is 2.0 $\pm$ 1.6 mm (0.0 to 5.1 mm) for GM variations. In the case of the tES, the maximum distance error is 2.0 $\pm$ 1.6 mm (0.0 to 5.4 mm) for CSF variations. We can observe that the maximum variations are within 5 mm. The detailed data are presented in Table~\ref{Tab1}.


\begin{figure}[htb]
\centering
\includegraphics[width=1.0\textwidth]{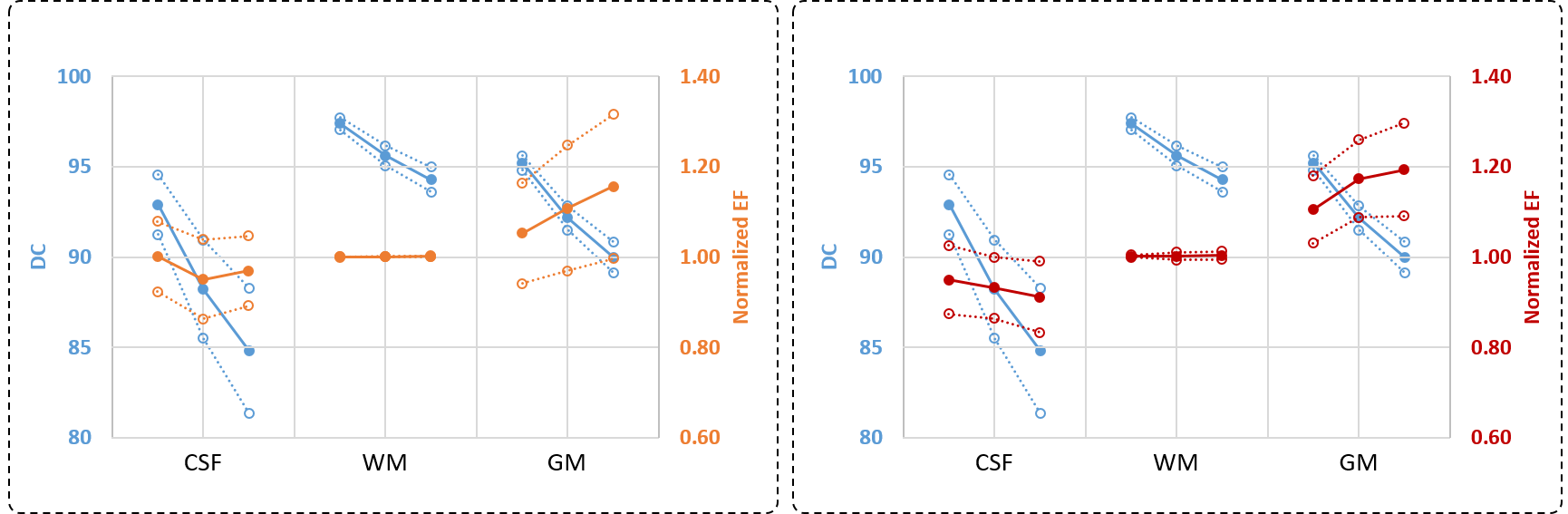}
\caption{Demonstration of how segmentation accuracy in terms of DC in CSF, WM and GM would influence normalized EF in TMS (left) and tES (20 $mm$ electrode) (right). Solid lines represent mean values and dashed lines represent mean$\pm$SD.}
\label{conclusionfig}
\end{figure}


\section{Discussion}

Deep learning is now emerging in different research fields and made significant improvements in terms of computation time, data quality, pattern understanding. In the near future, it is expected to be standard approach for different difficult tasks such as anatomical segmentation of medical images. However, it is still unclear how deep learning can be carefully optimized to provide a non-biased segmentation that can work efficiently in general form. A conventional problem in medical image segmentation is how to find the threshold value that demonstrate accurate segmentation from probabilistic maps. Common approach is to assume that all components are treated equally and high score wins even if difference is marginal. However, this is unfair decision considering different characteristics and patterns of image components. Moreover, sensitivity of segmentation accuracy is attributed to the clinical application. It is worth noting that our target here is to provide a guideline for users to select a segmentation method (or fine tune existing one) such that it carefully and accurately segments sensitive tissues. What are these tissues? How sensitive are they? And what is the cost of segmentation error? These are the questions that we are trying to discuss here. In this study, we investigated how parametric segmentation that was based on deep learning probabilistic maps can lead changes in electric field distribution for different brain stimulation applications. This will definitely help us to understand how brain stimulation computations is sensitive to variability of segmentation.

We considered ForkNet segmentation as it can successfully segment MRI images into 13 head tissues. Then, parametric segmentation was conducted by emphasizing a single-tissue a time that is demonstrate a more favor segmentation weights in regions with high uncertainties. Results demonstrated that different tissues behaved in different ways which can be refereed to tissue volume, tissue contrast in MR image, neighbor tissues and other related spatial properties. In experimental study, we consider segmentation using single MRI T1 only, which was proved to be enough for a reasonable deep learning segmentation \Bcite{Wachinger2018NI,Rashed2019NI,Roy2019NI}. Considering multimodality anatomical images may improve the segmentation quality, however, another factor (registration error) should be carefully analyzed for validation of segmentation error effect.  

A brain stimulation using TMS and tES are conducted considering variations in CSF, WM and GM. Summery of the obtained results are demonstrated in figure~\Bref{conclusionfig}. Results present an interesting insight. In both applications, changes corresponding to WM is found to produce a marginal change in the electric field distribution in the motor cortex. However, electric field may have a point-wise difference around 20\% as shown in the cross sectional error images in figure~\Bref{TMSerror}. It is clear that CSF and GM leads to notable change of electric field with higher $\omega$ values. However, a different behavior is recognized with CSF and GM. Increasing $\omega$ for CSF (GM) leads to decrease (increase) of the induced electric field in TMS and tES. Segmentation variation lead to volume change of the annotated tissues as well as surrounding ones. Therefore, a change in the induced electric field is expected with range identified by the target tissues and surrounding ones. Induced electric field show high sensitivity to CSF segmentation and neighboring tissues such as GM. This is clear from the reverse tendency between CSF and GM.

A potential limitation of this study is that the parametric segmentation is based on T1 MRI only. A more accurate results may result using multi-modality segmentation (including T2). Moreover, it is also interesting to extended the current study on deep brain regions.


\section{Conclusion}

Segmentation of different tissues is an important step in the standard pipeline in the head model development for electromagnetic simulation. However, segmentation is not an easy task due to different factors such as PVE. Also, it is still unclear how accurate segmentation is required for reliable computations of the distribution of the induced electric field. We present a method for parametric segmentation that can generate head models with different variations using tissue probability maps generated by deep learning architecture. We study the influence of segmentation error in each tissue and how it is correlated with the distribution of induced electric field. Results indicated that some tissues are of high sensitivity to segmentation errors such as CSF, while other tissues are less sensitive such as WM for head magnetic-exposure when investigating effects on brain cortex. This study focus on electric field within M1 region. The insights obtained in the present study are useful when considering other body parts in particular as systematic full-body is complicated at the moment.


\section*{Acknowledgment}

This work was supported in part by the Ministry of Internal Affairs and Communications, Japan. Grant Number JPMI10001.

\section*{References}
\bibliographystyle{dcu}
\bibliography{Refs}
\end{document}